\documentclass{article}

\usepackage[utf8]{inputenc} 
\usepackage[T1]{fontenc}    
\usepackage{hyperref}       
\usepackage{url}            
\usepackage{booktabs}       
\usepackage{amsfonts}       
\usepackage{amsmath}
\usepackage{amssymb}
\usepackage{nicefrac}       
\usepackage{microtype}      
\usepackage{graphicx}
\usepackage{svg}

\usepackage{xcolor}

\usepackage[backend=biber, sorting=none, style=phys]{biblatex}

\usepackage{mathtools}

\usepackage{caption}
\usepackage{subcaption}
\usepackage{fancyhdr}
\usepackage{fancybox}
\usepackage[verbose=true,letterpaper,top=1in,bottom=1in,left=0.75in,right=0.75in]{geometry}
\usepackage{authblk}
\usepackage{arxiv}

\usepackage[export]{adjustbox}
\def\eff{\mathrm{eff}}
\def\WT{w}
\def\MT{m}

\title{Geometry-Induced Competitive Release in a Meta-Population Model of Range Expansions in Disordered Environments}

\author[1]{Jimmy Gonzalez Nu\~nez}
\author[1]{Daniel A.~Beller}

\affil[1]{Department of Physics and Astronomy, Johns Hopkins University, Baltimore, MD, 21218}

\hypersetup{
    citebordercolor={1 1 1}
}

\begin{document}
\maketitle

\begin{abstract}
    Rare evolutionary events, such as the rise to prominence of deleterious mutations, can have drastic impacts on the evolution of growing populations. Heterogeneous environments may reduce the influence of selection on evolutionary outcomes through various mechanisms, including pinning of genetic lineages and of the population fronts. These effects play significant roles in enabling competitive release of otherwise trapped mutations. In this work we show that environments containing random arrangements of ``hotspot'' patches, where locally abundant resources enhance growth rates equally for all sub-populations, give rise to massively enriched deleterious mutant clones. We derive a geometrical optics description of mutant bubbles, which result from interactions with hotspots, that successfully predicts the observed increase in mutant survival. This prediction requires no fitting parameters and holds well in  scenarios of rare mutations and of adaptation from standing variation. In addition, we find that the influence of environmental noise in shaping the fate of rare mutations is maximal near a percolation transition of overlapping discs, beyond which mutant survival decreases. 
\end{abstract}


\twocolumn
\newpage

\section*{Introduction}\label{sec:intro}

In well-mixed microbial systems, the timing of spontaneous mutations can have significant consequences for their fate and impact on a population's evolution, with mutations arising early in the expansion history benefiting from prolonged growth \cite{Luria1943}. In contrast to populations grown in well-mixed conditions, spatially structured populations undergoing range expansion characteristically experience ``gene surfing'' of late-occurring mutations at the colony periphery, resulting in an enrichment of high-frequency clones that promotes rare evolutionary outcomes \cite{Hallatschek2008,Klopfstein2005,Hallatschek_2007,Edmonds2004,Excoffier2008,vanGestel2014,Nadell2013,Buttery2012}.

A signature of gene surfing during range expansions is the formation of regions, called ``genetic sectors'' \cite{Hallatschek_2007},  composed primarily of genetically identical individuals. Due to stochasticity in reproduction times, the boundaries of these sectors fluctuate laterally with increasing expansion distance. When a sector's boundaries coalesce to a point, the sector loses contact with the front and terminates as a ``bubble'' \cite{Fusco2016,Gralka2019} with characteristic length  $l_{\parallel}$ parallel to the expansion direction. For sufficiently slow diffusion of nutrients, growth occurs only very near the front, so terminated sectors are cut off from further growth and have no presence in the remainder of the range expansion.   In the absence of environmental stressors, such as antibiotics \cite{Haeno2007}, mutations that arise with a higher metabolic cost compared to the wild-type  population will have a decreased establishment probability (to form sectors) as they tend to lose contact with the front more rapidly, thus producing bubbles of smaller $l_\parallel$. 

Microbial range expansion studies \cite{Hallatschek_2007,,Borer2020,Korolev_2011,Hallatschek2008} have made substantial progress in characterizing  neutral evolution during a range expansion. Importantly, there is evidence that sector boundaries fluctuate superdiffusively with dynamics described by the Kardar-Parisi-Zhang (KPZ) universality class \cite{Kardar1986}. Further studies have found that KPZ dynamics also describe the scaling of mutant bubble sizes, a finding that holds not only in neutral evolution but also with rare, deleterious mutations \cite{Fusco2016}.

Environments can modify mutant bubble dynamics and the likelihood of deleterious mutants forming sectors (survival) rather than bubbles (extinction). For example, studies on heterogeneous landscapes of obstacles (regions of locally suppressed growth) have revealed that in such an environment, population fronts are pinned by the local heterogeneity, which causes an effective reduction in the selective advantage of wild-type sub-populations \cite{Gralka2019}. 
Furthermore, on landscapes of patches that locally allow only the deleterious mutant to grow, sufficiently close spacing of patches allows the mutant to survive through a phenomenon of  \textit{assisted percolation} \cite{tunstall2023assisted}. 
This ``competitive release'', which enhances the establishment probability of deleterious mutations, can also occur for time-varying environmental changes, such as a sudden introduction of antibiotics \cite{Fusco2016, Segre2016Quantifying}. Competitive release is related to evolutionary rescue, in which deleterious mutations increase in prominence as a result of environmental changes during population growth \cite{Fusco2016, Bell2009}.

 
Recent studies have shed light on the mechanisms by which interactions between population fronts and environmental structure  profoundly alter evolutionary trends. An individual obstacle acts similarly to a bump in surface topography by focusing the expansion front inward to a caustic \cite{Moebius2015,Beller2018}, and an individual hotspot acts as a radiation zone that propels an advancing front radially outward \cite{Moebius2021,nunez2024range}. In the context of neutral evolution, a disordered landscape of hotspots was recently shown to replace a population's intrinsic demographic noise by environmentally determined genetic structure at large scales, with successful genealogical lineages pinned to fastest paths through a subset of the hotspots \cite{nunez2024range}. However, little is known about the influence of such quenched-random noise on evolutionary processes involving mutation and selection. The characterization of how random spatial variation in high-nutrient regions influences the survival of deleterious mutations is a missing component of a broader understanding of how environments affect the outcomes of rare evolutionary events. 

In this work, we use a meta-population model based on the Eden model for range expansions, with growth coupled  to environmental structure through a distribution of hotspots, to study how environmental quenched-random noise modifies the survival of a deleterious mutant. Despite locally giving equal benefits in growth rate to the wild-type and the mutant, the hotspots nonetheless cause massive enrichment of deleterious mutant clones by reducing the effective selective advantage of the wild-type. We characterize the spatially non-uniform mechanism for mutant clone enrichment, which involves ``lanes'' of high mutant survival probability.  By constructing and iterating upon a geometrical model for the average influence of individual hotspots, we provide an effective description for the mutant survival probability found in our simulations. 

\section*{Meta-Population Model for Spatial Growth}

We implement a two-species version of the Eden model \cite{eden1961two}, coupled to a landscape of hotspots. This simple meta-population model has been extensively studied and used in investigations of evolutionary dynamics, including phenomena such as fitness collapse \cite{Lavrentovich2015a}, fixation \cite{Krishnan2019}, and gene surfing \cite{Paulose_2020,Gralka2016}, and it generates clone size distributions that reproduce experimental observations \cite{Fusco2016}.

\begin{figure}
    \centering
    \includegraphics[width=0.49\textwidth]{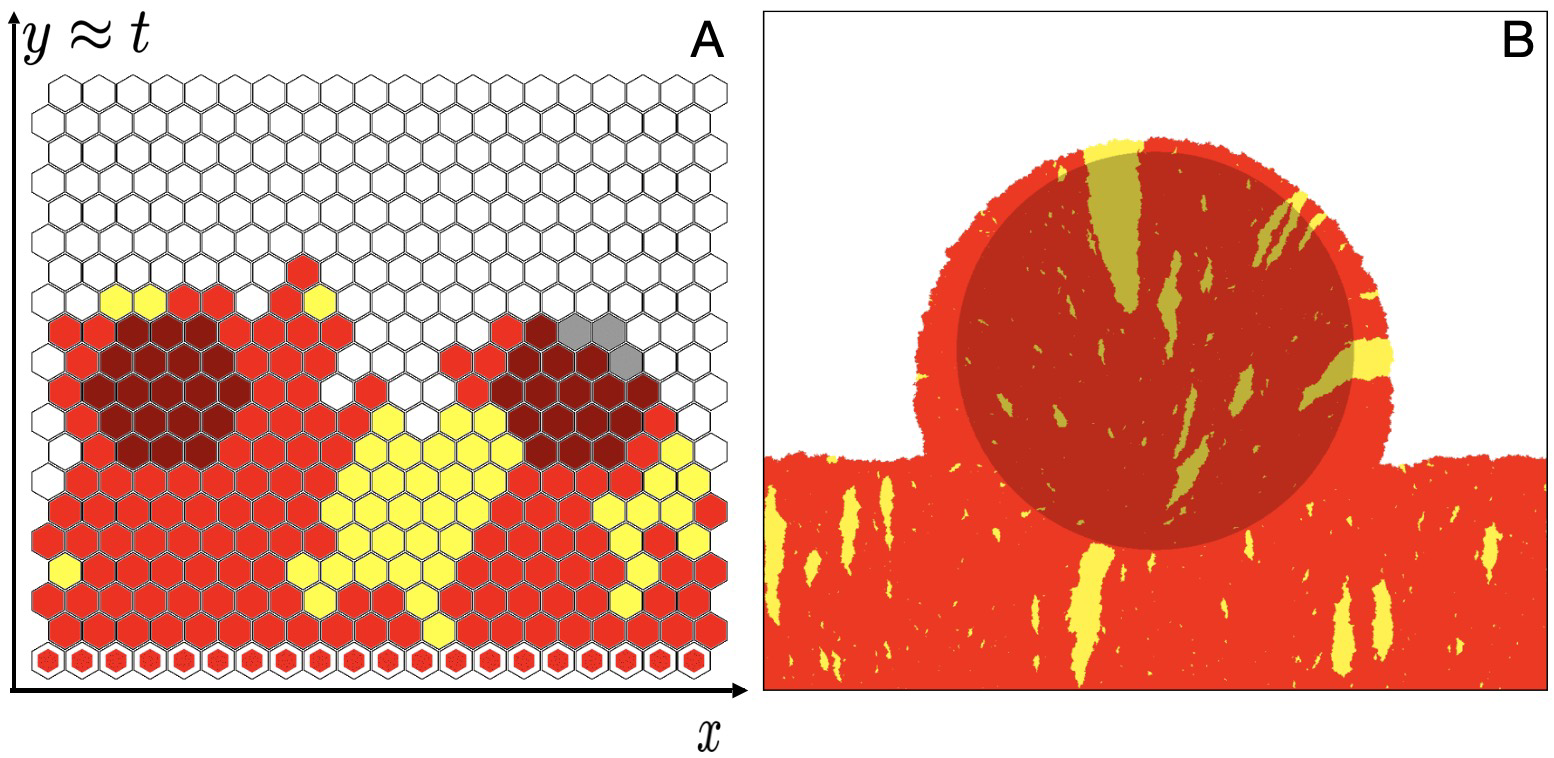}
    \caption{(A) Illustration of a two-species Eden model for a range expansion on a heterogeneous landscape of hotspots (darkener-shaded sites) with a linear initial condition (white-outlined hexagon markers) of wild-type (red) and mutants (yellow) arising at a mutation rate $\mu$. (B) Simulation snapshot showing the effect of a single hotspot on the population front and mutant bubbles (yellow). Snapshots were generated using simulation parameters: (A) $\nu=10, R=2, s =0.1, \mu=0.01$ and (B) $\nu=10, R= 100, s=0.01, \mu=0.01$.}
    \label{hexgraph}
\end{figure}

The two sub-populations in our model are a \emph{wild-type} and a \emph{mutant}, the latter having a selective disadvantage compared to the former. Our simulated population is arranged on a hexagonal grid, with each filled grid site containing a locally well-mixed  \emph{deme}, which can grow by filling an empty neighboring site (Fig.~\ref{hexgraph}A, \href{https://pages.jh.edu/dbeller3/resources/SI-Videos/Gonzalez-Nunez-arXiv-2024b/Movie S1.gif}{Supplementary Movie S1}). The genetic character of each deme is assumed to be uniquely determined by the first individual to arrive. This corresponds to the regime where growth to local carrying capacity occurs faster than migration or mutation. A single identifier thus characterizes the local genetic composition, which we represent with one of two distinct colors corresponding to wild-type (red) or mutants (yellow). 

    To model competition for scarce resources and space, our system allows  reproduction of demes only at the expansion front where the population borders the empty region; the dynamics behind the front is ``frozen'', with no changes allowed to occupied sites. Initial conditions consist of a single, filled line of $L$ sites at the bottom edge on the hexagonal grid. The mutant's growth rate $\Gamma_m = 1 - s$ is smaller than that of the wild-type, $\Gamma_w = 1$, by the wild-type's selective fitness advantage $s$. Deleterious mutations correspond to the interval $1 > s > 0$. When a wild-type deme replicates into a neighboring lattice site, the newly filled site is a mutant deme with probability $\mu$, the mutation rate, and wild-type with probability $1-\mu$. Back mutation of mutant to wild-type is assumed not to occur, so replication of a mutant deme always fills a neighboring empty site with another mutant deme. 

We implement quenched environmental noise in the form of hotspots, which are circular patches of increased replication rate that represent regions of higher nutrient availability. The hotspots have the same effect on both sub-populations, in the sense that they increase the replication rate of all demes inside them by the same constant multiplicative factor $(1+\nu)$. Here, the hotspot strength $\nu$ is a dimensionless parameter, with $1 + \nu = \Gamma^H_i/ \Gamma^B_i$ defined as the ratio of growth rates within a hotspot ($H$) and in the surrounding bulk ($B$), for both wild-type ($i=w$) and mutant ($i=m$).
At scales larger than the deme size, the emergent front propagation speed in the Eden model of each sub-population is directly proportional to the local growth rate, $v^X_i \propto \Gamma^X_i$ ($X \in \{H,B\}$). Thus, the hotspot intensity can also be defined as $\nu = v^H_i / v^B_i - 1$ ($i \in \{m, w\}$). A collection of randomly placed, possibly overlapping hotspots constitutes a landscape, characterized by hotspot radius $R$, hotspot intensity $\nu$, and hotspot area fraction $\phi$. An important length scale, the typical hotspot center-to-center distance, is given by \cite{Xia1988,Torquato2002}
\begin{equation}\label{eq:lambda_from_phi}
    \lambda(\phi, R) = \sqrt{2}R\sqrt{\frac{\pi}{-\ln(1 - \phi)}}.
\end{equation}

The two-species model is summarized by the following rules for replication rates $\Gamma^X_i$ within hotspots ($X=H$) and in the surrounding bulk ($X=B$), for mutant and wild-type demes:
\begin{align}
  \Gamma_w^{X}  & \overset{\mu}  {\rightarrow} \Gamma^X_m,\;\; X \in \{B, H\}  & \text{(mutation)}   \label{eq: mutation} \\ 
   \Gamma^X_m & = (1-s)\Gamma^X_{w}, \;\; X \in \{B, H\}  & \text{(selection)} \label{eq: selection} \\ 
  \Gamma_i^{{H}} &= (1 + \nu) \Gamma^B_i, \;\; i \in \{w, m\}  & \text{(hotspot strength)} \label{eq: speedup} 
\end{align}
The absence of back mutation means that there is no reverse process to Eq.~\ref{eq: mutation}, $\Gamma_m^X\rightarrow \Gamma_w^X$.

Reproduction is implemented using replication rules known to produce experimentally observed meandering statistics of sector boundaries \cite{Hallatschek_2007} and ancestral lineages \cite{Gralka2016}, associated with interface roughening of population fronts in the KPZ universality class \cite{Halpin_Healy_1995}. This asynchronous reproduction approach  fills empty sites one at a time according to the following procedure: (1) the population front is identified as the set of demes that are adjacent to at least one empty site, (2) one such population-front deme is randomly selected according to an implementation of the Gillespie algorithm \cite{Gillespie1976,Cai2009} that enforces Eqs.~\ref{eq: selection} and \ref{eq: speedup} on average, and (3) from among the chosen deme's empty neighbor-sites, one site is randomly selected with uniform probability to establish a new deme with inherited genotype, subject to the mutation rule of Eq.~\ref{eq: mutation}. Note that for neutral evolution in uniform environments, this model reduces to selecting individual demes at the population front with equal probability and copying their genotype (color) to a random adjacent empty grid site, which is the type C variant of the Eden model \cite{eden1961two,Jullien1985}.

\section*{Response of Population Structure  to Environmental Structure}

\begin{figure}
    \centering
    \includegraphics[width=0.49\textwidth]{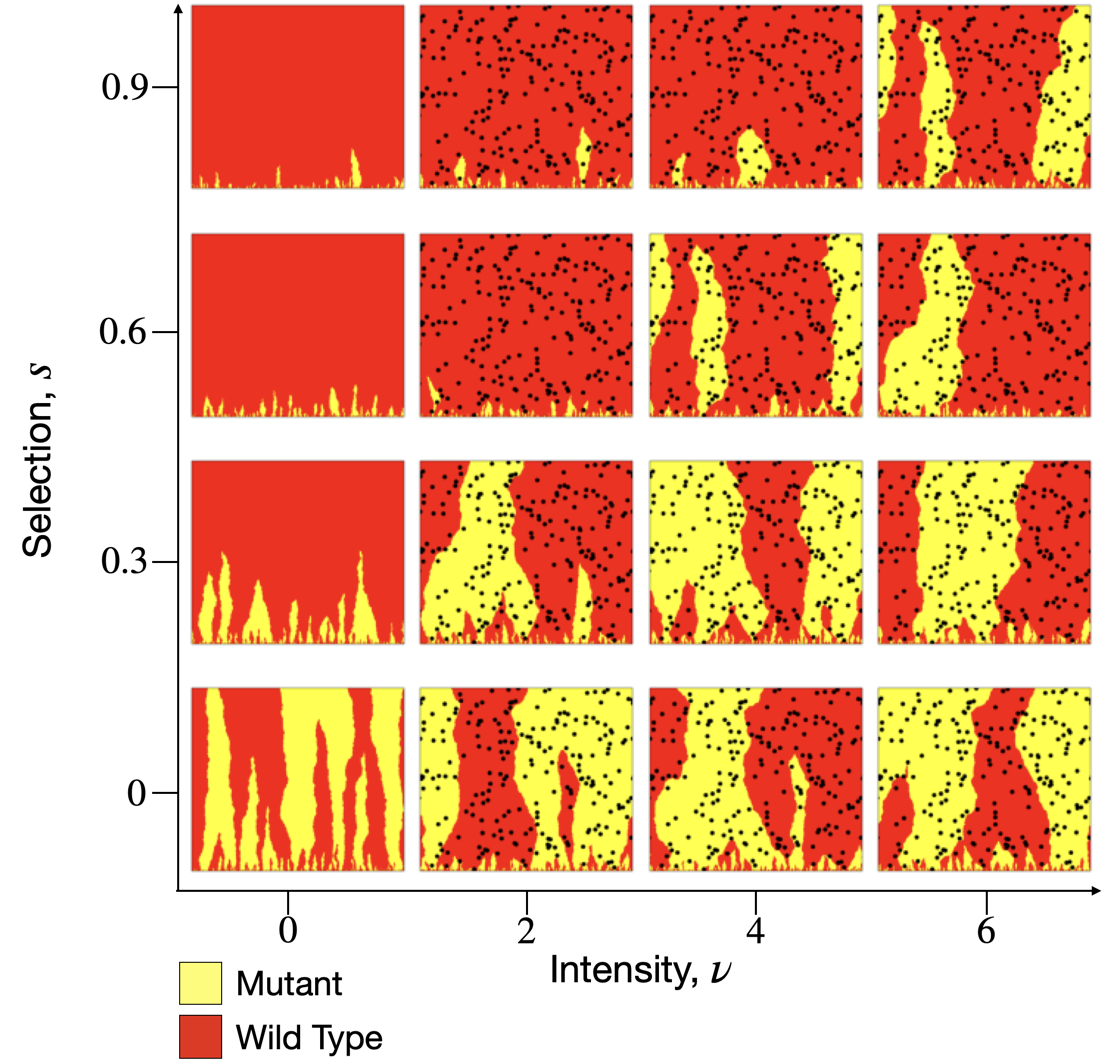}
    \caption{Simulation snapshots from a fixed landscape of hotspots (black discs) with radius $R=10$ and area fraction $\phi = 0.1$, using a standing-variation initial condition and zero mutation rate, for various hotspot intensities $\nu$ and selective advantages $s$.}
    \label{phase_diagram_snapshots}
\end{figure}

To understand how a disordered landscape of hotspots affects the fate of a deleterious mutation, we first examine the case of standing variation, taking $\mu=0$ but including mutants in the initial population. We construct the initial population as a row of  alternating wild-type and mutant sites, which minimizes initial correlation lengths and ensures that no sub-population has an increased chance of survival due to initial population sizes.

For a fixed distribution of hotspots of radius $R$ covering a fraction of the landscape area, $\phi$, the two governing system parameters are the hotspot intensity $\nu$ and the selective advantage $s$ of the wild-type. Snapshots of simulations from the same initial random seed are shown in Fig.~\ref{phase_diagram_snapshots}. In the absence of environmental noise, increasing $s$ reduces the lifetime $l_{||}$ of the mutant clonal domains. For non-zero selection, increasing the intensity of hotspots results in an increase in mutant survival likelihood, as  the largest mutant domains exhibit longer lifetimes. This trend is qualitatively similar to the effective reduction of selection by environmental noise in the form of randomly placed obstacles \cite{Gralka2019}. We will show in the next section that the reduction in selection efficacy is determined by the hotspot separation length scale $\lambda$ and the hotspot intensity $\nu$.

\begin{figure}[htb]
    \centering
    \includegraphics[width=0.48\textwidth]{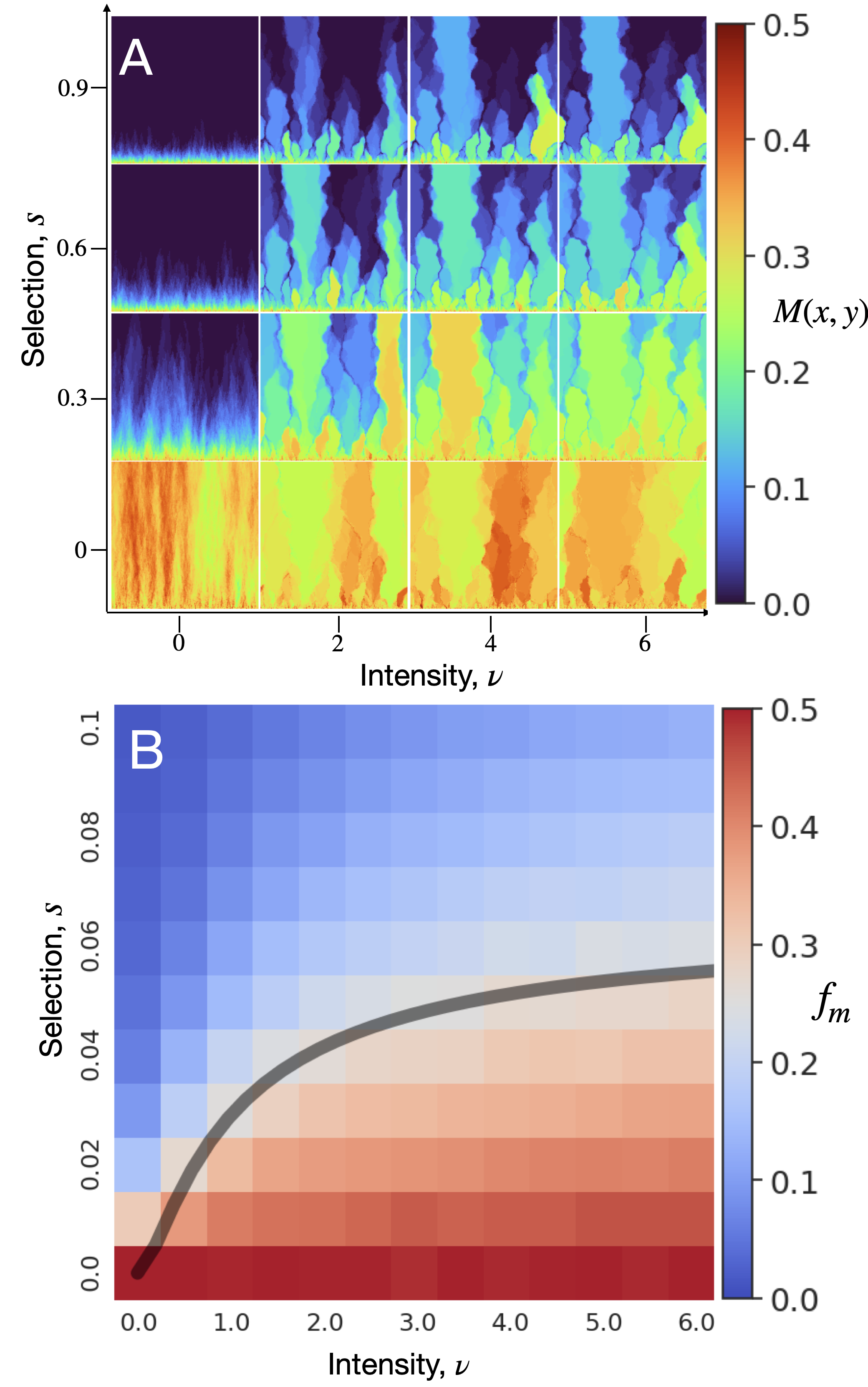}
    \caption{
    (A) Local mutant frequency $M(x,y)$,  on a fixed landscape of hotspots, averaged over different initial seeds. (B) Mutant frequency $f_{m}$ heatmap for combinations of selection, $s$, and hotspot intensity, $\nu$. Predicted critical selection $s_c$ (gray curve) from Eq.~\ref{eq:s_eff} provides a good approximation to the $f_m = 0.25$ contour. In both (A) and (B), the simulation ensemble is generated from 200 seeds per landscape and 20 landscapes for each pair of  $(s, \nu)$. Each simulation begins with an initial population consisting of 1000 sites of alternating wild-type and mutants, and each ends at a height of 1000 sites. The mutation rate $\mu$ is set to zero. The hotspot radius is $R=10$ and the hotspot area fraction is $\phi=0.25$.
  }
    \label{phase_diagram}
\end{figure}

To explore the conditions in which a disordered landscape enhances deleterious mutant survival, we study the average sector behavior by determining the probability $M(x,y)$ that a deme at position $(x,y)$ is the mutant type, computed from an ensemble of 200 independent simulations on a single landscape. Here and throughout, we call regions of mutant probability $M(x,y) > 0.5$  ``mutant domains''; these may be ``sector domains'' or ``bubble domains'' depending on whether they are connected to the front, analogously with sectors and bubbles in individual simulation runs.
$M(x,y)$ is shown in Fig.~\ref{phase_diagram}A for various combinations of $\nu$ and $s$ all on the same landscape, revealing a key feature: for $\nu > 0$, large mutant domains emerge and remain mostly fixed in position as  selection $s$ or intensity $\nu$ is increased. This observation hints that there are ``lanes'' through the landscape of hotspots that provide boosts in growth and thereby locally increase mutant survival likelihood. 

\begin{figure}[ht]
    \centering
    \includegraphics[width=1\textwidth]{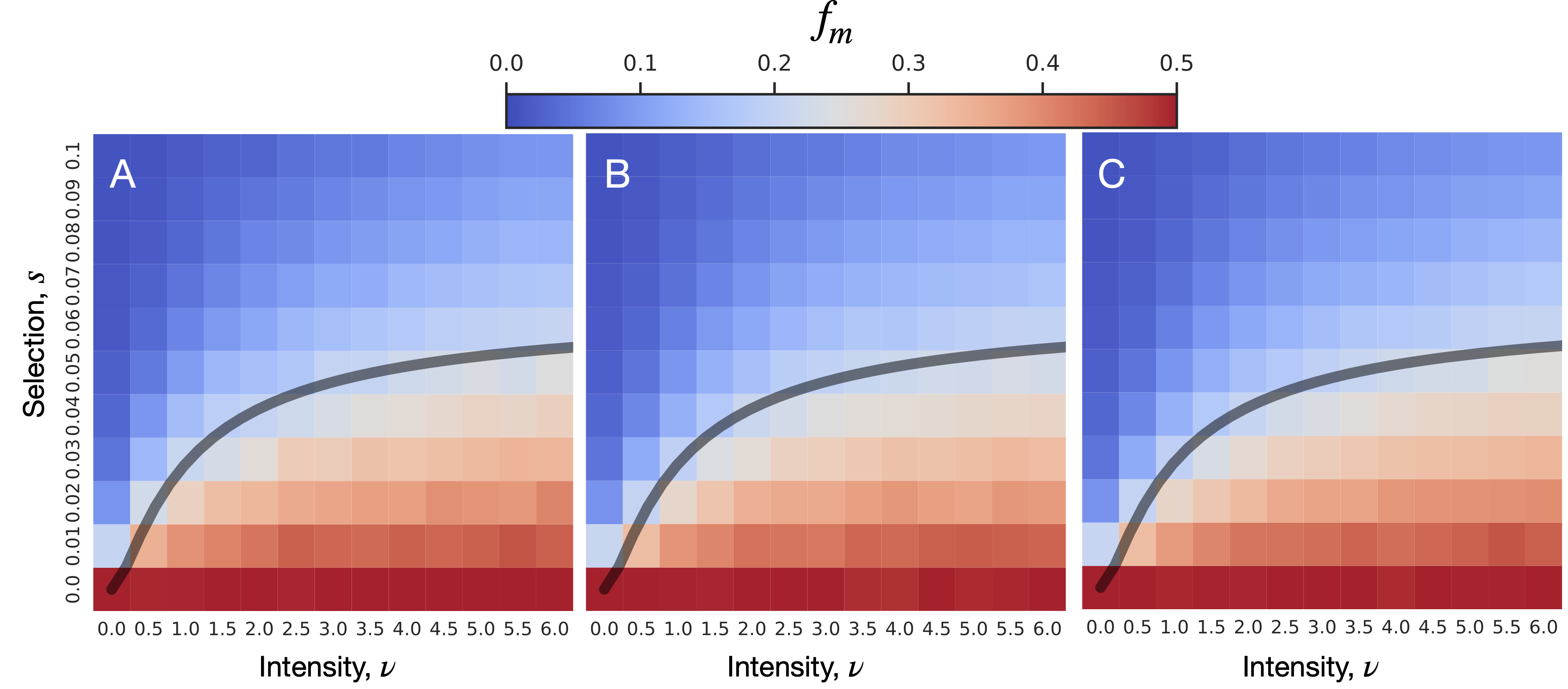}
    \caption{
    Mutant frequency $f_{m}$ heatmap for combinations of selection, $s$, and hotspot intensity, $\nu$. The critical selection $s_c$ according to Eq.~\ref{eq:s_eff} is plotted in gray and approximates the $f_m = 0.25$ contour. Simulation ensemble is generated from 200 seeds per landscape and 20 landscapes for each pair of  $(s, \nu)$. Each simulation begins with an initial population consisting of $L$ sites of alternating wild-type and mutants, and each ends at a height of $h$ sites. The mutation rate $\mu$ is set to zero. The hotspot radius is $R$ and the hotspot area fraction is $\phi=0.25$. (A) $L=2000$, $h=2000$, $R=20$. (B) $L=1000$, $h=2000$, $R=10$. (C) $L=2000$, $h=2000$, $R=10$.}
    \label{different_sizes}
\end{figure}

The survival trends for sectors over an ensemble of landscapes can be summarized by constructing a phase diagram for the mutant frequency, $f_{\MT} \equiv \langle M(x,y) \rangle$, determined as the mean ratio of the total mutant clone area to the colony size, which is a direct observable in experiments using fluorescence microscopy \cite{Gralka2019}. In the null case of zero selection and zero environmental noise, neutral mutants will, on average, have equal clonal sizes compared to the wild-type, and thus $f_{\MT} = 0.5$, whereas $f_{\MT}\approx 0$  represents mutants that are extinguished in one generation. Thus, $f_{\MT} = 0.25$ sets the midpoint, with regard to mutant frequency, between the limits of infinitely strong selection and neutral evolution. The $(s, \nu)$ phase diagram for $f_{\MT}$ is shown in Fig.~\ref{phase_diagram}B, which reveals that the mutant frequency increases with increasing hotspot intensity at fixed selection. Contours of constant mutant frequency have slopes in the $s-\nu$ plane that decrease as $\nu$ increases. Phase diagrams for various system sizes are shown in \textit{SI Appendix}, Fig.~\ref{different_sizes}, demonstrating that the $f_m = 0.25$ contour is independent of system height.

The enhancement of mutant frequency relies on the spatial structure of the hotspot landscape, not merely the existence of the hotspots. To see this, we construct a minimal mean-field description in which each sub-population has an effective growth rate calculated as its spatially averaged growth rate over \emph{any} landscape with hotspot area fraction $\phi$ and hotspot intensity $\nu$:
\begin{align}\label{eq_effective_growth}
  \Gamma_{\eff}^{\WT} &= (1 - \phi) + \phi(1+\nu), \\
  \Gamma_{\eff}^{\MT} &= 
  (1-s) \left((1-\phi) + \phi(1+\nu)\right). \nonumber
\end{align}
From these effective growth rates, we define a corresponding effective selection coefficient that characterizes the competition between wild-type and mutant populations, 
\begin{equation}
  s_{\eff} = \frac{\Gamma_{\eff}^{\WT} - \Gamma_{\eff}^{\MT}}{  (\Gamma_{\eff}^{\WT} +  \Gamma_{\eff}^{\MT}) / 2}.
\end{equation}

In terms of this selection coefficient, the time-evolution of the mutant fraction $\rho_m(t)$ and of the wild-type fraction $\rho_{\WT} (t) = 1 - \rho_{\MT} (t)$ are described by
\begin{equation}
  \frac{\partial\rho_{\WT}}{\partial t} = s_{\eff} \rho_{\WT} (1 - \rho_{\WT}) - \mu \rho_{\WT},
\end{equation}

where $\mu$ is the mutation rate and is set to zero for the case of standing variation. There is one non-trivial stable point corresponding to equality of the two effective growth rates in Eq.~\ref{eq_effective_growth}. However, since  $\Gamma_{\eff}^{\WT} > \Gamma_{\eff}^{\MT}$ for all positive $s$, the only phase boundary in $(s,\nu)$ space predicted by this naive mean-field description is at $s=0$, whereas all systems with $s>0$ have $\rho_w=1$, $\rho_m=0$ as their long-time limit. Because this mean-field result is inconsistent with the continuous range of mutant frequency values that we have calculated for our model (Fig.~\ref{phase_diagram}B), the enrichment of mutant clonal domains (and thus the possibility of mutant survival) must be a consequence of the environment's spatial structure. Figure~\ref{phase_diagram}A suggests that any particular landscape of hotspots contains favorable paths for the establishment of surviving mutant domains, a process whose geometry we explore in the next section.

\section*{Expansion and Contraction Bubble Geometry}
\begin{figure}
    \centering
    \includegraphics[width=0.49\textwidth]{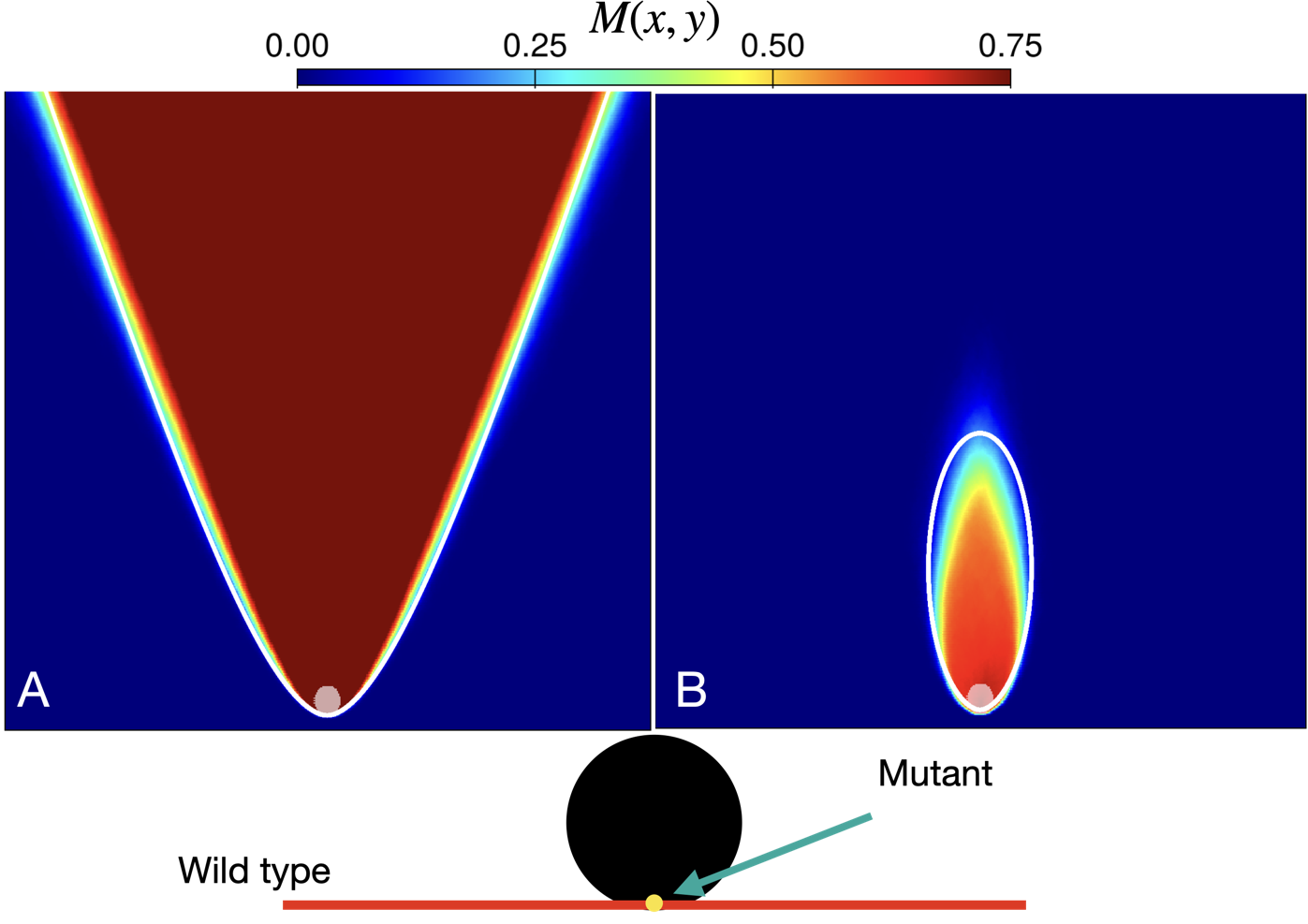}
    \caption{
    Mutant spatial frequency $M(x,y)$ for a population that is seeded with a single (A) beneficial mutant ($s=-0.1$) or (B) deleterious mutant ($s=0.15$), at the point where an all-wild-type front encounters the hotspot (bottom schematic). Heatmaps are generated from an ensemble of 1000 independent simulations with a single hotspot of radius $R=10$ and intensity $\nu=10$, on a grid of $500 \times 500$ sites.}
    \label{mutantPatterns}
\end{figure}

We gain insight into the influence of individual hotspots on the formation of mutant bubbles and sectors by examining the mutant domains of $M(x,y)$ for a population front composed entirely of wild-type demes except for a single mutant seeded at the point where the front encounters the hotspot, as shown in Fig.~\ref{mutantPatterns} (bottom schematic). We observe that a beneficial mutation ($s < 0$) forms a hyperbola-bounded sector domain (Fig.~\ref{mutantPatterns}A, white curve) with a high mutant frequency that is nearly uniform in the region above the hyperbola. On the other hand, a deleterious mutant ($s > 0$) forms an ellipse-bounded, flame-like bubble domain (Fig.~\ref{mutantPatterns}B, white curve) that tapers toward the end of its lifetime. 

\begin{figure}[t]
    \centering
    \includegraphics[width=0.48\textwidth]{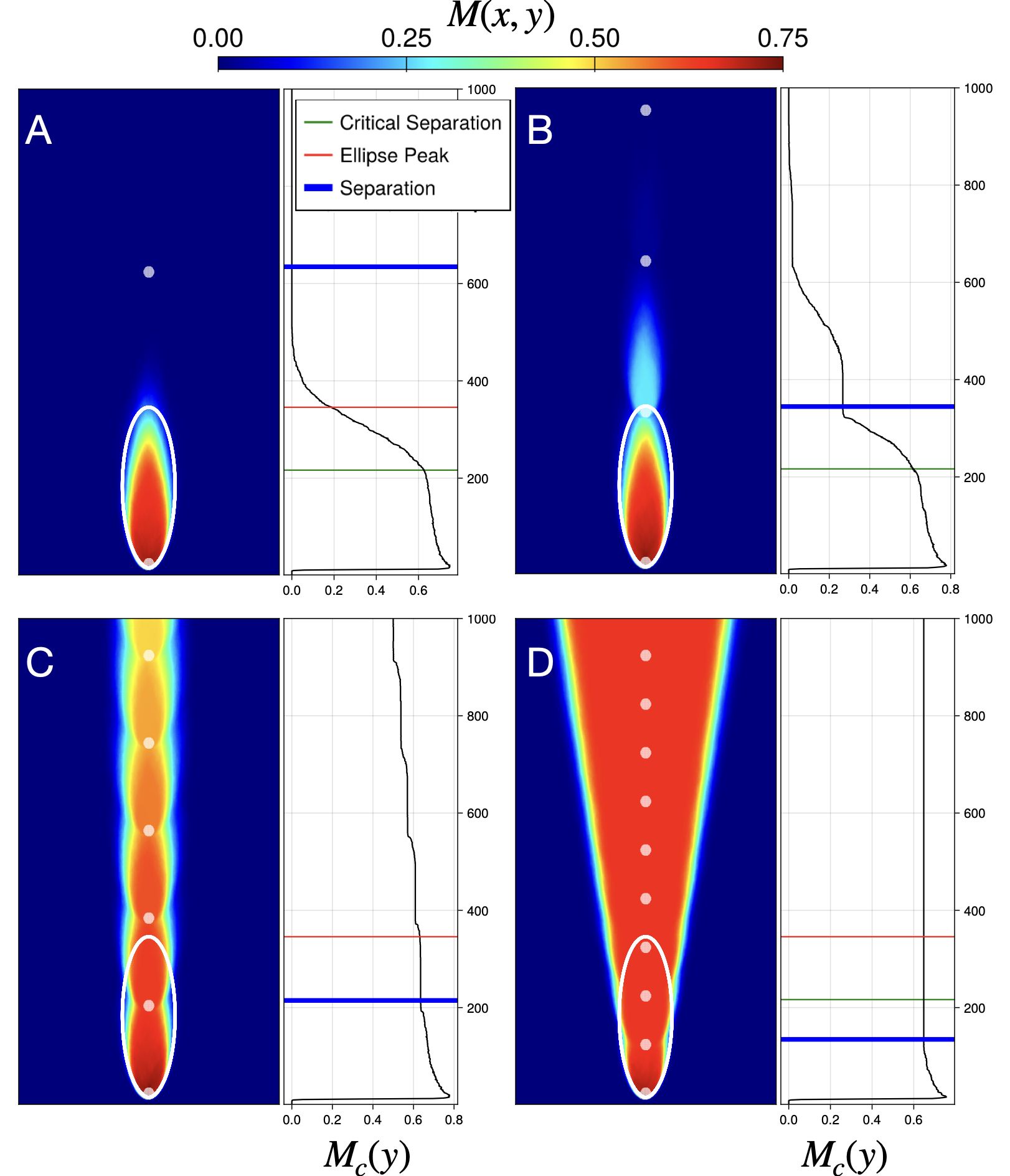}
    \caption{Mutant spatial frequency $M(x,y)$ for a population that is seeded with a single deleterious mutant ($s=0.1$) once a wild-type front encounters the first hotspot in a vertical sequence of hotspots separated by center-to-center distance $z$ for (A) $z = 600$, (B) $z = 310$, (C) $z = 180$, (D) $z = 100$. Heatmaps were generated from an ensemble of of 1000 independent simulations with hotspots of size $R=10$, hotspot intensity $\nu=10$, on a grid of $500 \times 1000$ sites. Right panels show the mutant spatial frequency along the center vertical line, $M_c(y) \equiv M(250, y)$. Also shown are reference horizontal lines corresponding to the hotspot position (blue), ellipse peak (red), and critical separation (green) of Eq.~\ref{eq:z_c}. }
    \label{extinction_expansion_transition}
\end{figure}

Following previously studied analogies between sector shapes induced by hotspots and geometrical optics \cite{nunez2024range, mobius2015obstacles,Moebius2021}, we apply the principle of least time to find the expected population front shape induced by passage through a hotspot. In particular, the expected (ensemble-averaged) front at some time is  the set of points that are first reached at that time by ``light rays'', originating normal to the initial population, that propagate with spatially dependent speed $v^H_i$, $v^B_i$ within and outside of hotspots, respectively. Likewise, the expected sector boundary of our single mutant sector in this scenario consists of the set of points that are reached in equal time by light rays passing through a hotspot and light rays outside of a hotspot. We simplify the calculation by considering very intense hotspots ($\nu \gg 1$). In this limit, determining light ray trajectories is equivalent to determining intersection points of the unperturbed population front and a circular population front originating at the hotspot center.  The resulting sector boundary is described, in polar coordinates, by the equation
\begin{equation}\label{eq:conic_sections}
  r(\theta) = \frac{2 R \sin(\theta)}{\gamma- \sin(\theta)},
\end{equation}
where $\gamma = 1 /[(\Gamma^B_{\WT} +  \Gamma^B_{\MT}) / 2] = 2 / (2 - s)$ is the inverse of the average population growth rate, and the origin is set at the bottom-most point of the hotspot. This equation predicts that the expected sector boundary traces out a hyperbola for $\gamma < 1$ $(s < 0)$ and an ellipse when $\gamma > 1$ $(s > 0)$, in agreement with our simulations (white curves in Fig.~\ref{mutantPatterns}A,B respectively). In the special case of zero selection ($s=0$), the resulting shape is a parabola, as has been seen in computational studies of neutral evolution \cite{nunez2024range,Moebius2021,Moebius2015}. 

For deleterious mutations ($s>0$), in order to understand the effects of these elliptical bubble domains on mutant survival in systems of many hotspots, we next consider a simplified landscape composed of a vertical sequence of hotspots. These hotspots are separated by center-to-center distance $z$, which we compare to the height $z_e$ of the ellipse from Eq.~\ref{eq:conic_sections}. Shown in Fig.~\ref{extinction_expansion_transition} is the spatial mutant frequency, $M(x,y)$, for various separation lengths $z$, with the same initial conditions employed in  Fig.~\ref{mutantPatterns}. We observe in Fig.~\ref{extinction_expansion_transition}A that mutant domains rapidly extinguish when hotspots are well-separated ($z \gg z_e$) due to a low probability of mutants reaching the next hotspot before losing contact with the front. As the hotspot separation decreases (Fig.~\ref{extinction_expansion_transition}B, $z \approx z_e$) the single mutant domain obtains an  increased lifetime (larger height) as mutants reach subsequent hotspots, but with decreasing probability at each successive hotspot, ultimately extinguishing at large expansion distances. When the hotspots are near a critical separation ($z \approx z_e / 2$) the mutant domain bubble transitions into a mutant sector domain, with the geometry of a vertical lane that reaches the final-time front. 
At smaller hotspot separations ($z < z_e /2$), the mutant domain boundary becomes a hyperbolic conic section, with nearly uniform $M(x,y)$ except at the outermost regions of the sector. 
Thus, repeated encounters with hotspots can enable luckily positioned deleterious  mutants to survive and even grow, with morphology similar to \emph{beneficial} mutations. We note a qualitative similarity of this finding to a key result of Ref.~\cite{tunstall2023assisted} for patches that prevent growth of the wild-type: The survival of a deleterious mutant requires sufficiently small spacing between patches in comparison to the height of a mutant bubble, whose shape the authors approximated as elliptical. 

An analytical solution for least-time trajectories is intractable for the scenario of many hotspots. Instead, we describe the average sector boundaries by constructing a minimal geometrical optics model. The simplification rests on the observation that the sector boundaries between mutant and wild-type in individual simulations are, up to model-dependent noise, given by the relative front propagation speeds of each sub-population. In our controlled scenario of a single mutation arising in a landscape of a vertical sequence of hotspots (Fig.~\ref{extinction_expansion_transition}), the mutant sub-population will be the first to receive a boost from the initial hotspot. Assuming that the boost is large enough for the mutant to outcompete the wild-type near the hotspot, the mutant sub-population will be the main beneficiary of sequential boosts as the front passes through each hotspot. We can then define an effective mutant front speed as the average speed of the tip of the population front,
\begin{equation}\label{eq:v_m}
  \bar{v}_m = \frac{v_m z}{z - 2 R + 2 R (1 + \nu)^{-1}},
\end{equation}
where $v_m $ is the (bare) mutant front propagation speed in the bulk. In this picture, a crossover from contraction (effectively deleterious) to expansion (effectively beneficial) selection for the mutant is predicted to happen when $\bar{v}_m$ is comparable to the (bare) wild-type front speed in the bulk, $\bar{v}_m = v_w \equiv 1$. In terms of the hotspot separation $z$, this transition occurs  when $z$ equals a crossover hotspot separation length given by
\begin{equation}\label{eq:z_c}
  z_c = \frac{2 R \nu}{(1 + \nu)s}.
\end{equation} 

An animation of this transition as $z$ passes through $z_c$ is shown in \href{https://pages.jh.edu/dbeller3/resources/SI-Videos/Gonzalez-Nunez-arXiv-2024b/Movie S2.gif}{Supplementary Movie S2}, demonstrating the transition from elliptical to hyperbolic conic sections. Additionally, $\bar v_m$ determines the angle $\theta$ formed by sector boundaries with the expansion direction, as $\tan (\theta)$ equals the ratio of the difference in sub-population front speeds to the mean front speed, $\tan(\theta) = 2(\bar{v}_m - v_w) / (v_w + v_m)$. Note that this relation produces vertical average sector boundaries, $\theta = 0$, in the case of neutral evolution, $\bar{v}_m = v_w$. The sector boundary angles are shown in \href{https://pages.jh.edu/dbeller3/resources/SI-Videos/Gonzalez-Nunez-arXiv-2024b/Movie S2.gif}{Supplementary Movie S2} (green lines), confirming our determination of the effective mutant sub-population front speed $\bar{v}_m$ in Eq.~\ref{eq:v_m}.

What lessons does this contrived scenario hold for mutant survival in the disordered landscape of many, randomly distributed hotspots? We conjecture that selection will be significantly suppressed when the typical hotspot separation length $\lambda$, given in Eq.~\ref{eq:lambda_from_phi}, for the disordered landscape is less than the $\nu$-dependent crossover hotspot separation $z_c$ calculated in Eq.~\ref{eq:z_c} for the vertical line of hotspots. 
By inverting the relationship in Eq.~\ref{eq:z_c}, we define a critical selection, $s_{c}$, as the value of the wild-type selective advantage $s$ that would produce a constant-width sector domain in a landscape with a vertical line of hotspots of separation $z$. We then replace $z$ with $\lambda$, by our conjecture, and use Eq.~\ref{eq:lambda_from_phi} to write $s_c$ in terms of the hotspot size, intensity, and area fraction
\begin{equation}\label{eq:s_eff}
  s_c(\nu, R, \phi) = \frac{2 R^2}{k\lambda^2} \frac{\nu^2}{(1 + \nu)^2}.
\end{equation}
Here we have used the fact that the transverse speed is proportional to $\sqrt{2k s}$ \cite{Hallatschek2010} with $k = 2\sqrt{2} / \sqrt{3}$ on a hexagonal lattice. The relation given by Eq.~\ref{eq:s_eff} is plotted in Fig.~\ref{phase_diagram}B (gray curve). With no fitting parameters, this form predicts the transition between strong and fully suppressed selection well by approximating the $f_m=0.25$ contour. 

\begin{figure}[h]
    \centering
    \includegraphics[width=0.45\textwidth]{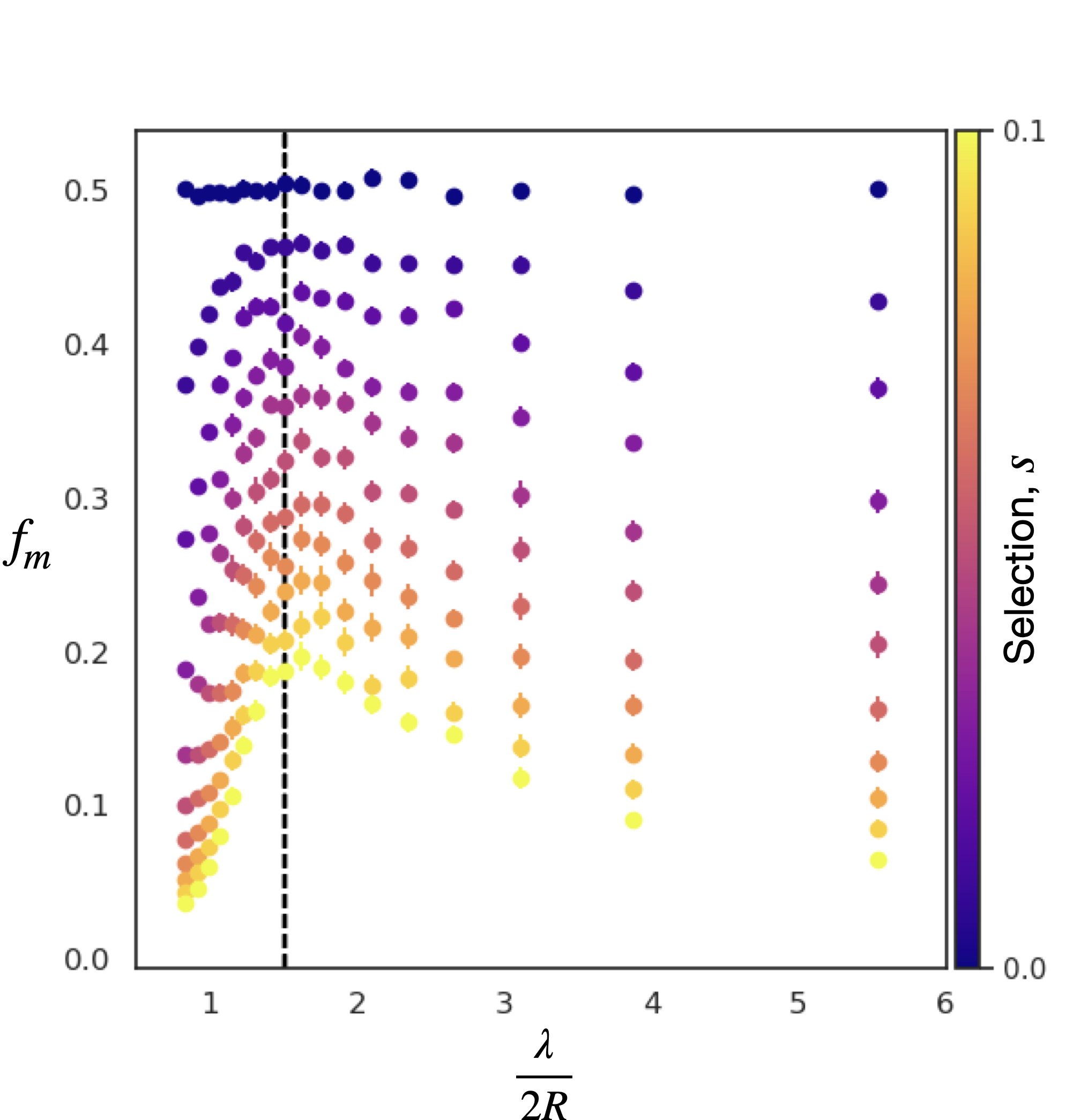}
    \caption{Mutant frequency $f_{\MT}$ as a function of the normalized hotspot separation length $\lambda / 2R$ for selection $s \in [0,0.1]$, with  $\nu=10, R = 10, \mu=0$. Vertical dashed line corresponds to an area fraction $\phi = 0.54$ and $\lambda / (2R) = 1.4$. Uncertainty bars represent one  standard error of the mean.}
    \label{fm_vs_lambda}
\end{figure}
We now examine how the mutant frequency depends on hotspot separation $\lambda$. 
We choose a high hotspot intensity ($\nu = 10$) to ensure that the first individuals reaching a hotspot receive a sufficiently large boost to outcompete neighboring demes entering a hotspot. Shown in Fig.~\ref{fm_vs_lambda} is the mutation frequency $f_{\MT}$ averaged across 20 landscapes, each with 200 simulations, as a function of $\lambda$ scaled by the hotspot diameter $2R$. As expected, $f_{\MT}$ decreases for all $\lambda$ as the selection $s$ increases because higher selection generally reduces  mutant bubble sizes. This effect of decreasing $f_{\MT}$ can be understood by noting that  $z_c / 2R \sim s^{-1}$ (Eq.~\ref{eq:z_c}), and thus larger selection $s$ requires closer spacing between hotspots in order for competitive release of mutants to be probable. Interestingly, Fig.~\ref{fm_vs_lambda} shows a non-monotonic dependence of $f_\MT$ on $\lambda$ with a maximum at $\lambda / (2R) \approx 1.4$.
For values of $\lambda$ that are small compared to $2R$, there is a high degree of hotspot overlap, and the landscape approaches a nearly uniform landscape (dominated by the higher growth rate of the hotspots), a situation that favors the wild-type by construction. In terms of hotspot area fraction (Eq.~\ref{eq:lambda_from_phi}), $f_{\MT}$ is maximal near an area fraction of 
$\phi \approx 0.54$. 

\section*{Clone Size Distributions in Heterogeneous Environments}
\begin{figure}[b]
    \centering
    \includegraphics[width=0.49\textwidth]{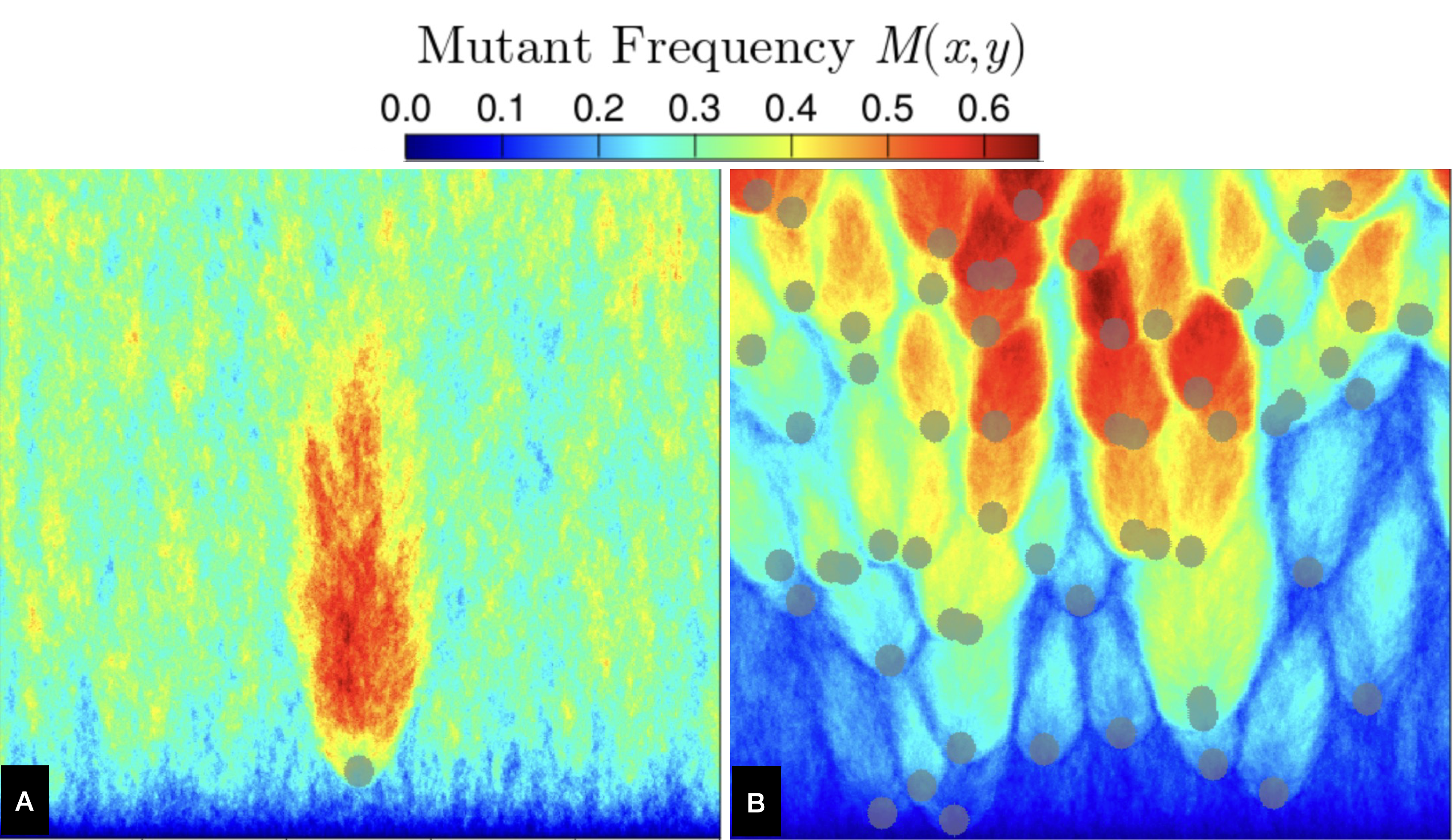}
    \caption{Mutant spatial frequency $M(x,y)$ for an initially all-wild-type population with mutation rate $\mu = 10^{-2}$, using 1000 independent simulations on the same hotspot landscape (gray disks) for (A) a single hotspot and (B) many randomly placed hotspots at area fraction $\phi=0.1$. In both (A) and (B), hotspots have radius $R=10$ and the landscape consists of $500 \times 500$ sites.}
    \label{close_ups_mutant_frequencies}
\end{figure}

We have seen in Fig.~\ref{extinction_expansion_transition} and the previous section that rare mutations that are fortunate enough to survive by gene surfing until encountering a hotspot will locally outcompete nearby wild-type sectors, increasing the probability of mutant survival at long times. 
In this section we show that our geometrical optics description for the extinction-survival transition from standing variation can also predict an extinction-survival transition for rare mutations that arise with constant probability per replication $\mu$. 

The mutant spatial frequency $M(x,y)$ for an all-wild-type initial population with mutation rate $\mu=10^{-2}$ is shown Fig.~\ref{close_ups_mutant_frequencies}. We observe that a single hotspot (Fig.~\ref{close_ups_mutant_frequencies}A) produces a mutant domain with a flame-like structure emanating from the hotspot, similar to Fig.~\ref{mutantPatterns}B. Unlike Fig.~\ref{mutantPatterns}B, Fig.~\ref{close_ups_mutant_frequencies}A has a nonzero background value of $M(x,y)$ because the nonzero mutation rate creates mutant bubbles at random locations; away from hotspots, these mutant bubbles generally have short lifetimes due to selection. A caveat in comparison to Fig.~\ref{mutantPatterns}B is that, with back-mutations disallowed,  mutants will necessarily dominate the population front for very large expansion distances \cite{Kuhr2011}. 

The flame-like mutant domains found in Fig.~\ref{close_ups_mutant_frequencies}A  also emerge in disordered landscapes of many hotspots with nonzero mutation rate, as shown in Fig.~\ref{close_ups_mutant_frequencies}B. In contrast to Fig.~\ref{close_ups_mutant_frequencies}A, there is a decreased probability of finding mutants in regions between the mutant domain bubbles. In regions where these domain bubbles overlap to form lanes through the landscape, mutants acquire an increased survival probability at large expansion distances.

\begin{figure}
    \centering
    \includegraphics[width=0.49\textwidth]{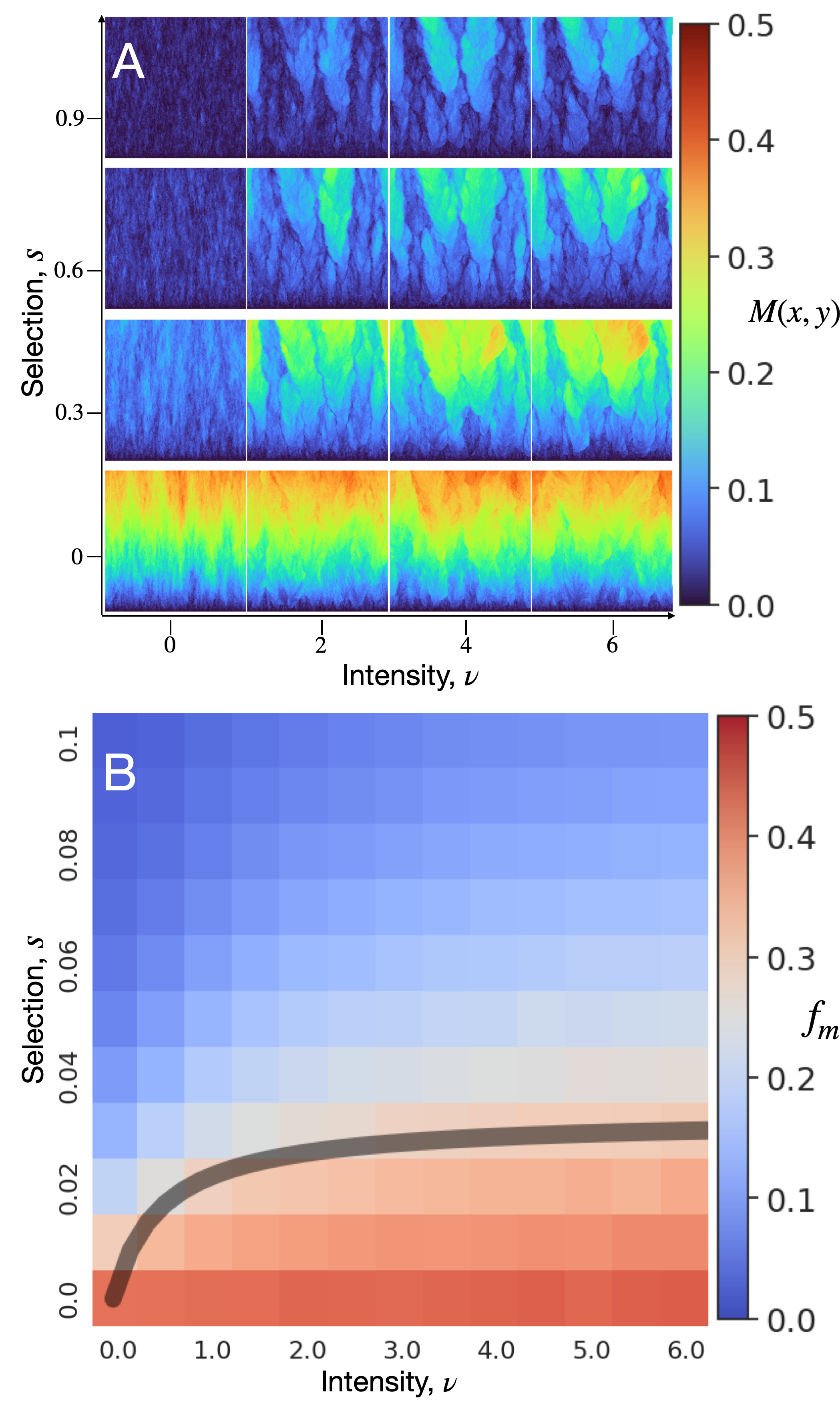}
    \caption{
    (A) Mutant spatial frequency $M(x,y)$ for fixed landscape of hotspots of area fraction $\phi=0.1$ at 
     hotspot intensities $\nu$ and selective advantages $s$. (B) Mutant frequency $f_\MT$, with phase boundary prediction shown as gray curve corresponding to $f_\MT =0.25$. 
     In both (A) and (B), the initial population consists of 1000 sites of entirely wild-type demes and the simulation ends at a height of 1000 sites. The ensemble is generated from 200 independent simulations for each of 20 distinct landscapes for each pair of  $(s, \nu)$. The mutation rate is set to $\mu=1\times 10^{-3}$, hotspot radius to $R=10$, and area fraction to $\phi=0.1$.}
    \label{phase_diagram_mutations}
\end{figure}

Since mutant bubble domains generated by a single hotspot in Fig.~\ref{close_ups_mutant_frequencies}A share many similarities with those in Fig.~\ref{mutantPatterns}B, we expect that our geometrical optics description (Eq.~\ref{eq:s_eff}) ought to also predict the extinction-survival transition for scenarios with nonzero mutation rate. In Fig.~\ref{phase_diagram_mutations}, we repeat the $M(x,y)$ and $f_\MT$ plots of Fig.~\ref{phase_diagram} but now using an all wild-type initial condition and $\mu=10^{-3}$. We note two qualitative ways in which the population structure in this constant-mutation-rate scenario (Fig~\ref{phase_diagram_mutations}) differs from the standing variation scenario (Fig.~\ref{phase_diagram}): first,  for neutral evolution ($s=0$),  $M(x,y)$ increases with expansion distance; second, mutant lanes tend to increase in number, rather than decrease, with expansion distance. Despite these notable differences in $M(x,y)$, we find that the phase diagram for$f_\MT$ (Fig~\ref{phase_diagram_mutations}B) shows a $f_m=0.25$ contour in ($s,\nu$) parameter space similar to that of Fig~\ref{phase_diagram}B and likewise well-described by Eq.~\ref{eq:s_eff}.

\begin{figure*}
    \centering
    \includegraphics[width=0.8\textwidth]{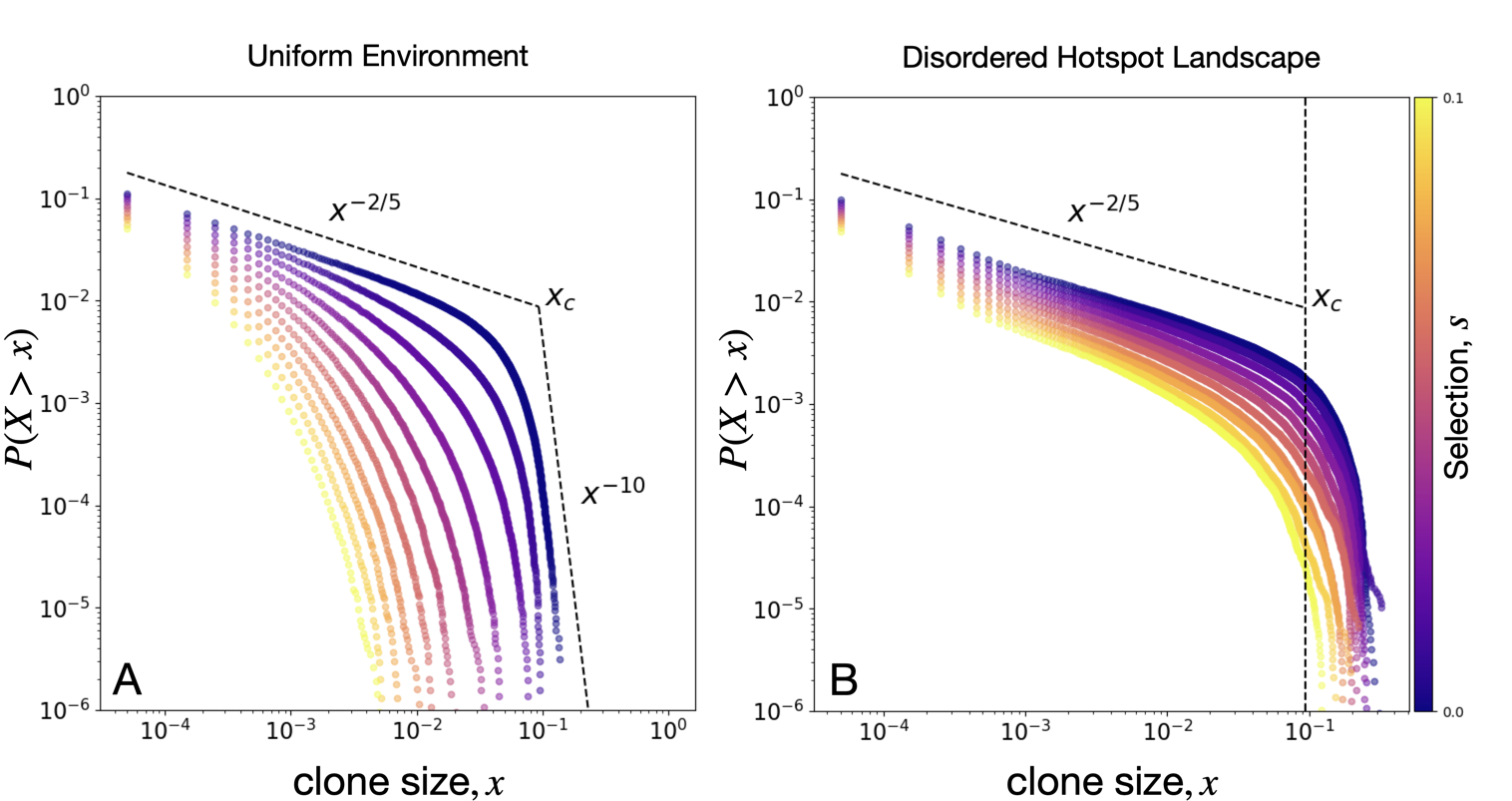}
    \caption{
    Mutant clone size distribution $P(X > x)$ for (A) a uniform environment and (B) a disordered landscape of hotspots with $R=10$, $\phi=0.1$, and $\nu=6$. Simulation ensemble consists of 20 landscapes each with 200 independent simulations for each combination of $\nu, s$. Dashed lines correspond to displayed power laws for neutral evolution in a uniform landscape, with $x_c$ being the expected largest mutant bubble.}
    \label{clone_distriubtion}
\end{figure*}

To make contact with experiments on evolution in microbial communities \cite{Fusco2016,Gralka2019}, we characterize the distribution of mutant clone sizes (area of mutant bubbles and sectors) in a disordered landscape of hotspots. This clone size distribution is directly related to the distribution in the number of single nucleotide polymorphisms (SNPs) \cite{Fusco2016} and to the site frequency spectrum measure used in population genetics to predict rare evolutionary outcomes, e.g.\ fitness valley crossings \cite{Weissman2009}. Specifically, we examine the reverse cumulative distribution $P(X > x)$ of clone sizes, which describes the probability that a random, rare mutation will produce a mutant bubble or sector of size (area) $x$ or larger. Shown in Fig.~\ref{clone_distriubtion} is $P(X > x)$ for $s \in [0,1]$. 

For a uniform landscape ($\nu = 0$, Fig.~\ref{clone_distriubtion}A), $P(X > x)$ exhibits a power law regime associated with the scaling of mutant bubbles at small clone sizes ($P(X >x)\sim x^{-\alpha}, \alpha=2/5$), 
The power-law scaling for bubbles in uniform landscapes is expected to hold for clone sizes smaller than $x_c = N^{-(1-\alpha)/(\beta - \alpha)} = 10^{-1}$, where $N=10^6$ is the number of lattice sites in our simulation and $\beta=4$ for KPZ growth processes \cite{Fusco2016}. The corner clone-size value $x_c$ separating these two regimes represents the largest emerging mutant bubbles in a uniform landscape with zero selection, and characterizes the transition from bubble scaling to sector scaling. Selection suppresses both mutant bubbles and sectors; as such, there is a reduction in $P(X > x)$ with increasing $s$, as shown in Fig.~\ref{clone_distriubtion}A. For example, the probability of finding mutant regions larger than $x_c$ vanishes for $s \geq 0.02$, indicating that mutant sectors are very unlikely to form, and the distribution of sector sizes $P(X>x)$ scales as $x^{-10}$ at large clone sizes; see \textit{SI Appendix} for details. 

\begin{figure}
    \centering
    \includegraphics[width=0.49\textwidth]{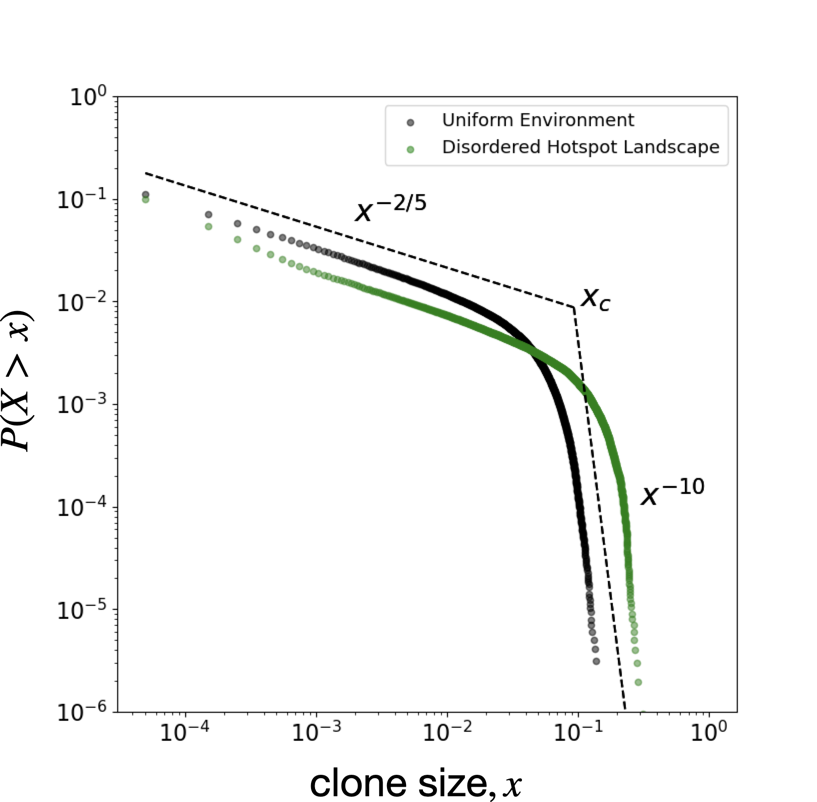}
    \caption{
    Mutant clone size distribution $P(X > x)$ for a uniform environment (black) and for a disordered landscape of hotspots (green) with $R=10$, $\phi=0.1$, and $\nu=6$. For both scenarios, simulation were performed using 20 landscapes each with 200 independent simulations, and with neutral evolution. Dashed lines correspond to displayed power laws for neutral evolution in a uniform landscape, with $x_c$ being the expected largest mutant bubble.}
    \label{clone_distriubtion_comp}
\end{figure}

The reduction in selection efficacy caused by a disordered landscape of hotspots promotes the formation of sectors and of larger bubbles, as shown in Fig.~\ref{clone_distriubtion}B. This enrichment of mutant clones is quantified by the non-zero likelihood of forming clone sizes  larger than $x_c$ for all $s$ values tested. However, the clone-size distribution for our studied range of $s$ retains the $P(X >x)\sim x^{-2/5}$ scaling that described neutral mutant bubbles in a uniform landscape. 
We also find that size distributions of deleterious mutant bubbles in disordered hotspot landscapes have similar scaling to size distributions of neutral mutant bubbles in uniform landscapes; see \textit{SI Appendix}, Fig.~\ref{clone_distriubtion_comp}. 
A possible explanation for this similarity in bubble scaling between environmental conditions is that, at low hotspot area fraction ($\phi= 0.1$), mutants emerge outside of a hotspot and form bubbles on scales smaller than the hotspot separation $\lambda$ where competition should resemble that in uniform landscapes. On the other hand, mutant bubbles that grow to sizes comparable to the hotspot separation, for example as shown in Fig.~\ref{extinction_expansion_transition}, become enriched by the landscape, generating formation of lanes and an excess of large clones $x > x_c$.


\section*{Conclusions}
Environmental structure can act as an extrinsic source of noise in a population's genetic structure, which tends to reduce selection efficacy and enhance the emergence of deleterious mutant sectors. In uniform environments, these deleterious mutations can remain trapped as bubbles in the non-growing bulk of the colony, behind the front. The trapped mutants are competitively released in environments that preferentially favor mutants \cite{tunstall2023assisted}, such as time-varying environments of antibiotics \cite{Fusco2016}. As we have shown in this work, competitive release of deleterious mutants can even occur in environments with features that benefit both wild-type and mutants. Thus, understanding the structure of clonal domains under multiple types of environmental heterogeneity is important for predicting the evolution of desired or undesired mutations.

Our findings indicate that environmental stressors selecting for specific phenotypes, such as application of drugs, or environmental features that suppress growth, such as landscape obstacles, are not necessary conditions for competitive release: Environmental heterogeneity that enhances reproduction of both sub-populations equally can similarly give rise to competitive release dynamics, with distinct spatial patterns of lanes of overlapping highly enriched regions. In both the standing variation and the rare mutation evolutionary contexts, we saw that environmental noise length-scales determine the statistics of mutant prevalence at the expansion front.

Despite the complex nature of gene surfing of rare mutations, a simple geometrical picture emerges in our findings for mutant clonal sectors in landscapes of hotspots: Individual hotspots produce one of three distinct spatial mutant clone patterns that are well-described by conic sections. These geometrical patterns, in combination with a simple geometrical optics analogy for propagating fronts, are sufficient to understand the main features of the complex mutant spatial structure in disordered landscapes of hotspots. Moreover, our results show that there exist optimal environmental conditions that maximize the survival of deleterious mutants initially present in a population. For our context of disk-shaped hotspots, this maximal chance of survival occurs near an area fraction $\phi=0.54$ at which typical hotspots have a very slight but nonzero overlap with their nearest neighbors, suggestive of a percolation transition. However, if the overlap between hotspots becomes too large, the prevalence of mutants in a population rapidly decreases. 

We have found that disordered environmental structure with no intrinsic advantages for a deleterious mutation can nonetheless induce competitive release of that mutation. By controlling landscape structure, it may be possible to engineer a mosaic of mutant bubble patterns that give rise to an enrichment of desired mutations,  conceivably including deleterious mutations to induce a mutational meltdown \cite{Castillo2020,LanschJusten2022,Jensen2020} where a population's growth is arrested by the overabundance of costly mutations. Tests of our predictions for mutant survival can be performed through a realization of hotspots in microbial range expansion experiments, with mutant clone size distributions and mutant frequencies both obtainable through fluorescence microscopy~\cite{Gralka2019, Fusco2016}. While we have focused on disk-shaped hotspots in this work for simplicity, our findings highlight the importance of understanding how more general disordered environments  may favor the survival of deleterious  mutations, and possibly hinder the spread of advantageous mutations. Our hotspot landscapes revealed that mutant lanes running through fastest paths \cite{nunez2024range} are key to deleterious mutant survival, a mechanism that future studies could use to understand evolutionary trends in disordered landscapes characterized by other forms of quenched-random noise.

\section*{Acknowledgements} 
The authors thank Jayson Paulose for many helpful discussions and Wolfram M\"obius for insights on hotspot landscapes. 

\printbibliography

\onecolumn
\section*{SI Appendix}

\begin{figure}[hb]
    \centering
    \includegraphics[width=0.95\textwidth]{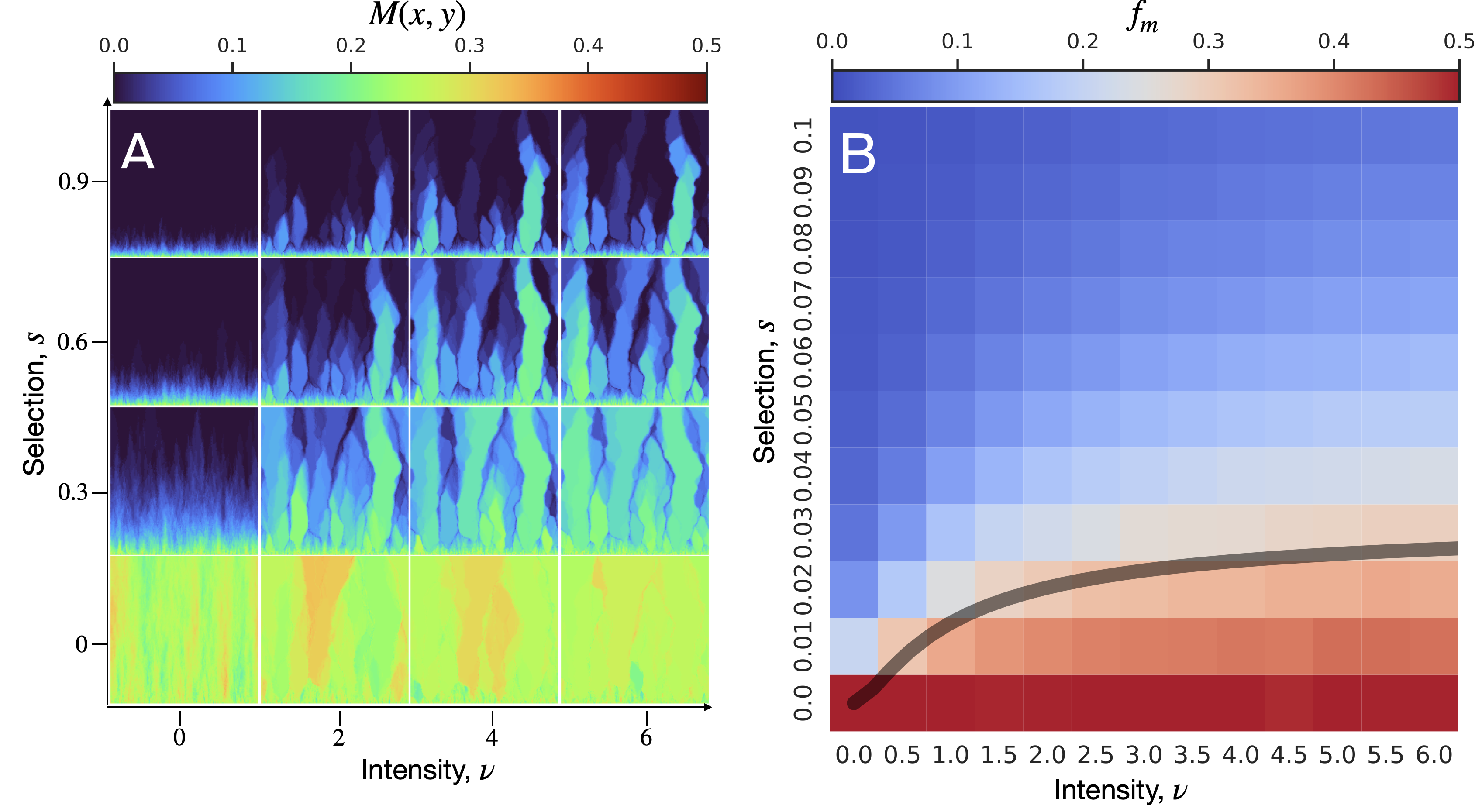}
    \caption{
     (A) Local mutant frequency $M(x,y)$,  on a fixed landscape of hotspots, averaged over different initial seeds. (B) Mutant frequency $f_{m}$ heatmap for combinations of selection, $s$, and hotspot intensity, $\nu$. Predicted critical selection $s_c$ (gray curve) from Eq.~\ref{eq:s_eff} and corresponding to the $f_m = 0.25$ contour. In both (A,B), simulation ensemble is generated from 200 seeds per landscape and 20 landscapes for each pair of  $(s, \nu)$. Each simulation begins from an initial population consisting of 1000 sites of alternating wild-type and mutants, and each ends at a height of 1000 sites. The mutation rate $\mu$ is set to zero. The hotspot radius is $R=10$ and the hotspot area fraction is $\phi=0.1$.
    }
    \label{fig:phase_diagram_0p1}
\end{figure}

\subsection*{KPZ Exponents in the Clone Size Distribution}
The exponents shown in Fig.~\ref{clone_distriubtion} can be derived for uniform landscapes from a direct mapping between superdiffusive random walkers and mutant bubble boundaries. Investigations using microbial colonies as model systems have shown that the dynamical scaling of sector boundaries in uniform landscapes is well described by the Kardar-Parisi-Zhang (KPZ) universality class \cite{Gralka2019,Fusco2016,Hallatschek_2007, Kardar1986}. In particular, the scalings of the lateral and parallel  components (with respect to the colony expansion direction) of bubble sizes are related by $L_{||} \sim L_{\perp}^z$, where $z$ is a dynamical exponent equal to  $3/2$ for the KPZ universality class. This means that mutant clone areas scale as $A \sim L_\perp^{1+z}$; equivalently, $A^{\frac{1}{1 + z}} \sim L_\perp $. For radial expansions, the clone size distribution exhibits by the following scaling relations, as shown in Ref.~\cite{Fusco2016}:  $P(A > a) = P(L_\perp > a^{\frac{1}{1 + z}}) \sim a^{\frac{-1}{1 + z}} = a^{-2/5}$ for  bubbles, and  $P(A > a) \sim  a ^{(1-2z)/ (z - 1)} = a^{-4}$ for sectors in radial expansion scenarios. In our system of linear expansion, which removes the inflationary effects that dominate radial expansion at late times, we find empirically that the asymptotic scaling of mutant sector areas is $P(A > a) \sim  a ^{-10}$.

\textbf{\href{https://pages.jh.edu/dbeller3/resources/SI-Videos/Gonzalez-Nunez-arXiv-2024b/Movie\%20S1.gif}{Supplementary Movie S1}}: Simulated range expansion on a hexagonal grid, with an initial population of wild-type (red) sites. Mutations (yellow) arise with a constant probability per replication.  Hotspots are shown as darker-shaded sites.

\textbf{\href{https://pages.jh.edu/dbeller3/resources/SI-Videos/Gonzalez-Nunez-arXiv-2024b/Movie\%20S2.gif}{Supplementary Movie S2}}: Mutant spatial frequency $M(x,y)$ for a range expansion across a landscape with a linear arrangement of hotspots (white disks). Elliptical sector boundary formed from a single hotspot is shown as white ellipse. Hotspots are separated by a fixed distance in each frame. The predicted sector angle is shown as a green line. Right-hand panel displays the mutant frequency along the center-vertical line. Position of the second hotspot, position of the peak of the ellipse, and critical separation are shown as blue, red, and green lines, respectively.

\end{document}